\documentclass[journal = apchd5,manuscript=article]{achemso}
\usepackage{amsmath}
\usepackage{xcolor}%
\usepackage{graphicx}
\usepackage{dcolumn}
\usepackage{bm}
\usepackage{subfigure}
\usepackage{textcomp}
\usepackage{natbib}
\usepackage{float}
\usepackage{xkeyval}
\bibpunct{}{}{,}{s}{}{;}
\usepackage{setspace}
\usepackage{caption}
\usepackage{geometry}
\usepackage[final]{pdfpages}

\newcommand{\ket}[1]{\left| #1 \right\rangle}
\newcommand{\bra}[1]{\left\langle #1 \right|}

\newcommand{\abs}[1]{\left| #1 \right|}

\newcommand{\expn}[1]{{\rm e}^{#1}}

\newcommand{\dg}{{^{\dagger}}}

\newcommand{\eg}{\textit{e.g.,}~}
\newcommand{\ie}{\textit{i.e.,}}
\newcommand{\nn}{\nonumber}

\SectionNumbersOff

\author{Steve M. Young}
\author{Mohan Sarovar}
\author{Fran\c{c}ois L\'{e}onard}
\email{*fleonar@sandia.gov}
\affiliation{Sandia National Laboratories, Livermore, CA, 94551, USA}

\title{Design of High-Performance Photon Number Resolving Photodetectors Based on Coherently Interacting Nanoscale Elements}

\begin{document}

\begin{tocentry}
\includegraphics[height=3.7cm]{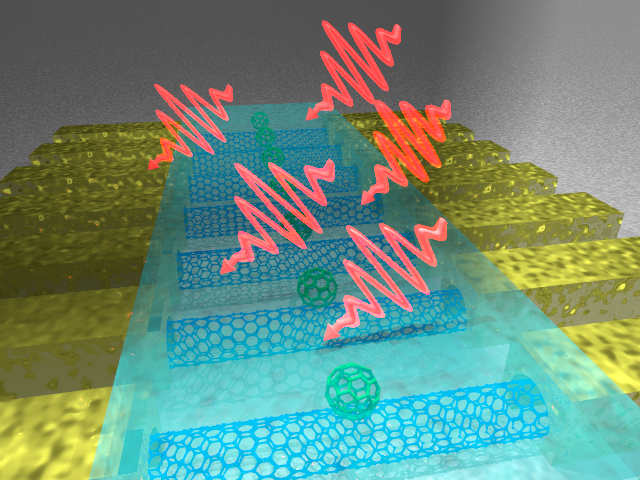}\\
	\textbf{Design for photon number resolving detector.} A new theoretical framework is used to design
	from the ground up a novel photodetector that achieves high performance and number resolution. The
	detector harnesses coherent interactions between nanoscale device elements, and could be realized with
	molecular systems.
\end{tocentry}
	
\begin{abstract}
A number of applications in basic science and technology would benefit from high fidelity photon number resolving photodetectors. While some recent experimental progress has been made in this direction, the requirements for true photon number resolution are stringent, and no design currently exists that achieves this goal. Here we employ techniques from fundamental quantum optics to demonstrate that detectors composed of subwavelength elements interacting collectively with the photon field can achieve high performance photon number resolution. We propose a new design that simultaneously achieves photon number resolution, high efficiency, low jitter, low dark counts, and high count rate. We discuss specific systems that satisfy the design requirements, pointing to the important role of nanoscale device elements.
\end{abstract}


\section{Introduction}
Detection and characterization of very weak light signals is an important capability in a number of scientific and engineering disciplines.  Significant progress has been made at the single photon level in photodetectors based on superconductors and avalanche photodiodes, the performance of which has achieved impressive levels across a number of important metrics \cite{Bienfang:2004ij,Woodson:2016cx,Pernice:2012bc,Marsili:2012ib,Marsili:2013th,Eisaman:2011cc, Hadfield:2009}.  However, building high-fidelity photon number resolving (PNR) detectors based on these platforms is challenging.  In both cases, the detector state is binary, so that detection can only indicate whether or not photons have been absorbed without information about the number present in the signal.  As a workaround, some degree of number resolution capability is achieved by multiplexing individual detectors into arrays \cite{Dicochiy:2008,Zhou:14,Ma_2017}.  While these arrays are capable of recording multiple photons there are significant drawbacks to this approach.  One is that nothing prevents an individual element from absorbing additional, unrecorded photons, resulting in reduced number resolution performance.  Second, such an array will necessarily occupy a greater area constraining its potential application space.

As an alternative approach, individual devices such as superconducting transition edge sensors \cite{Rosenberg:2005, Fukuda_2011} and semiconductor-based systems \cite{Kardinal:2008, Waks_2003} were demonstrated to resolve a few photons without multiplexing due to their ability to output a signal proportional to the number of photons in the detector. Such devices have already enabled some important
developments in quantum optics \cite{stobiska}. However, the non-unity efficiency due to the inherent difficulties in capturing and detecting all incident photons causes a reduction in the fidely of PNR.  Additionally, some of these detectors are more correctly described as energy-resolving detectors; number resolution therefore assumes monochromatic light with frequency known beforehand. 

From a theoretical perspective \cite{sperling_true_2012, Young:2018b, Jonsson:2018}, recent work has shown that high fidelity PNR requires detectors with very high efficiency for single-photon detection, and that the number of detectors required in a multiplexed approach increases rapidly with the number of photons to be resolved even for perfect efficiency detectors. While improvements in materials and fabrication may lead to performance improvements for the single photon detectors that make up these arrays, the above fundamental
limit is a general property of such systems. It is therefore important to consider other approaches that
could circumvent this fundamental limitation.

In addition to number resolution, there are several other metrics characterizing few-photon photodetector performance, including efficiency, dark count rate, jitter, and bandwidth. (Current PNR detectors
have not simultaneously attained high PNR fidelity and high performance in these metrics \cite{Lamas-Linares2013,Korzh2018}.)
Ideally, one desires a photodetector that excels at these metrics while also accurately enumerating the photons present in an incoming signal.  This means that the detector system must be capable of taking multiple distinguishable states corresponding to different numbers of absorbed photons and producing a unique output corresponding to each \cite{Dowling_2017}. However, while this is necessary, recent work has shown \cite{Young:2018b} that it is not sufficient; a simple array of ideal single-photon detectors satisfying this condition will nonetheless experience degraded performance in resolving photon number.
  Thus, in order to achieve efficient number resolution the entire detector must be modeled, designed and optimized as a complete and integrated whole, taking advantage of collective coherent interactions between all the detector elements and the optical field. We perform this task here, producing a general photodetector architecture 
  capable of efficient photon resolution while still meeting all other previously mentioned performance objectives.  We calculate the minimum model parameters required to achieve this performance, and discuss candidate physical platforms for implementing this architecture. In particular, we show that nanoscale device elements play an important role in the realization of these novel
detector architectures.

A novelty of the work presented here is the consideration of detectors composed of subwavelength/nanoscale detection elements that collectively and coherently interact with the photon field. Understanding
the photodetection performance of such a system requires going beyond standard photodetection theory
\cite{glauber_1963} to more advanced modeling and simulation tools that make fewer assumptions and treat the photon field, light-matter interaction, and transduction process as one coupled dynamical system.

\section{Description of photodetector}

\begin{figure}
\centering
\includegraphics[width=0.6\columnwidth]{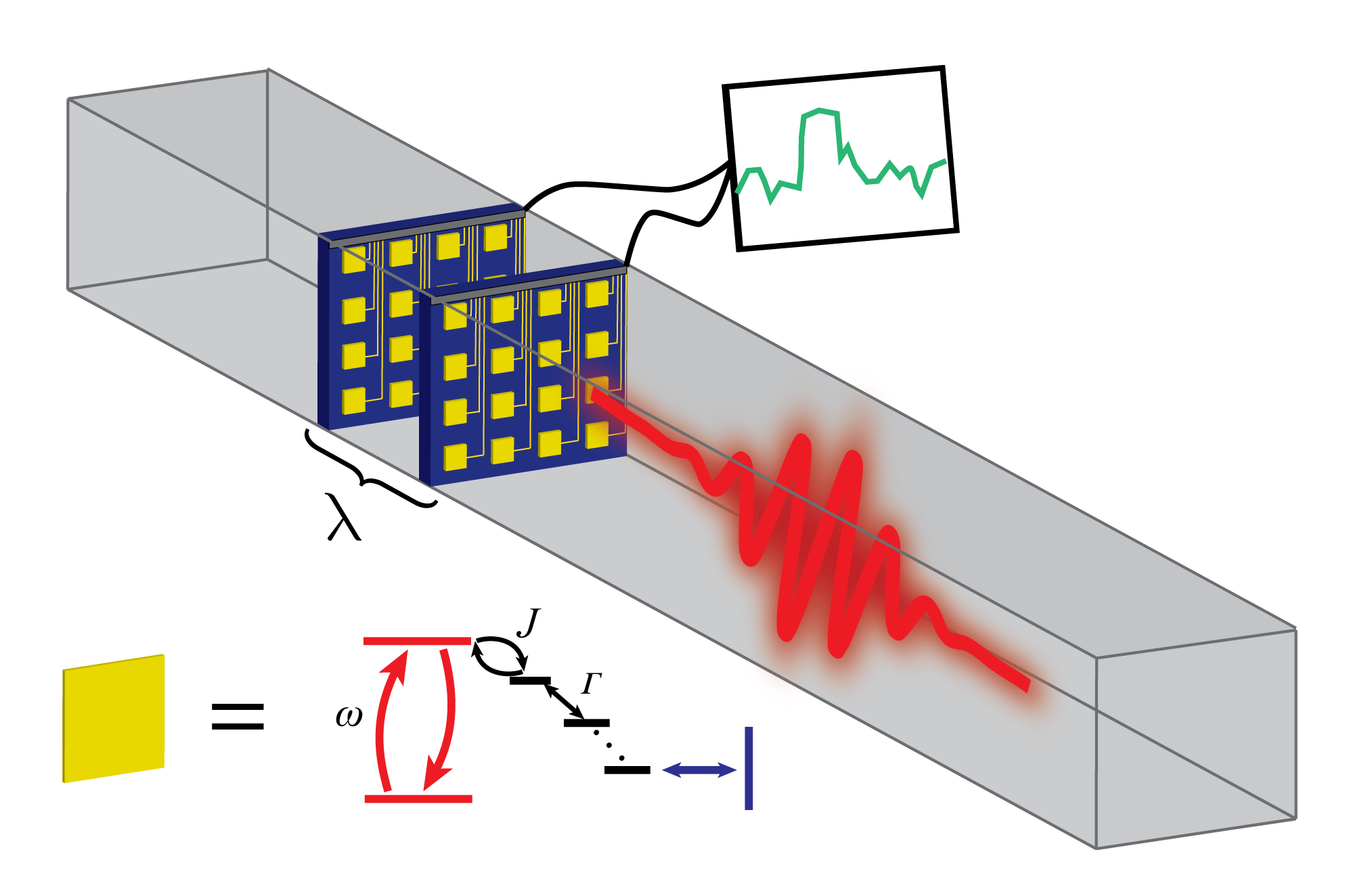}
\caption{ Illustration of the photon number resolving detector under consideration. A light pulse
containing multiple photons propagates in a waveguide. The detector is composed of individual
subwavelength elements represented by the yellow squares, each interacting with the photon field and generating a signal
when a single photon is absorbed. In general, an element can be described as having an absorption 
process (red), a transduction of the excitation through a number of coherent and incoherent
processes (black),  and measurement of the matter system to generate a signal (blue).
 All of the elements interact collectively and coherently with the field.  \label{fig:detector1}}
\end{figure}

The general photodetector architecture is illustrated in Figure~\ref{fig:detector1}. There, a photon pulse containing from 1 to $N$ photons is propagating in a single-mode waveguide. The detector itself is composed of a large
number of subwavelength/nanoscale detection elements organized in three dimensions, with all the elements in
the volume within the photon wavelength. Each of the elements is able to absorb photons and
to generate an electrical signal that is fed to processing electronics. In this work, we are concerned with
the operation of the photodetector after light delivery and up to the processing electronics, \ie~up to the generation of the electrical signal but not its further processing. Losses due to these additional steps will
affect the overall system performance, and will eventually need to be considered in order to compare with
existing approaches on an equal footing. We note that exquisite control over device fabrication and light delivery has already lead to minimization of these losses in existing platforms \cite{Lita:08}.
Intrinsic losses in the part that we model could also be present and are discussed below.

The various elements of the detector will together compose an optical absorption system, a transduction system that
transforms the optical excitation into other excitation modes (localized to the matter subsystem), and a measurement system that creates a signal based on monitoring of the internal modes. For example, in a superconducting nanowire
photodetector element, the absorption process corresponds to optical absorption by the superconducting
nanowire, the transduction process is the conversion of the excitation into local heating that destroys the
superconducting state, while the measurement process is the probing of the superconducting/normal state by an electrical current. A similar correspondence can be made to avalanche photodiodes where the avalanche
breakdown is the
transduction process. Other types of detection systems may also be mapped into this general scheme, such
as nanoscale electronic transport channels functionalized with absorbers \cite{Takesue}.

Because all of the detector elements in Figure~\ref{fig:detector1} are within the photon wavelength, they coherently and collectively
interact with the incoming photon field. In addition, we treat the photon field, the
matter subsystem, and the measurement process as part of one quantum system. This means that all
relevant processes are coupled; for example, the measurement process affects the matter subsystem,
which in turn impacts the photon field, and vice-versa. 

Despite the described device detecting light of a single spatial mode, it is nonetheless possible to efficiently detect photons in arbitrary spatial modes using a photonic lantern \cite{Birks:15}.  These devices were originally invented in the context of astronomy precisely for the purpose of demultiplexing multimode signals into many single modes.  The performance and function of such devices is being continuously improved as they find new applications \cite{Amado:18}.

\section{Theoretical framework}
Previously we developed a framework for modeling quantum photodetectors using a master equation approach \cite{Young:2018b}, including a quantized field constructed within the quantum noise framework and amplification/measurement processes treated according to continuous measurement theory. In this section, we summarize the key expressions and relations of this framework.

Our formalism allows one to evaluate the state of the matter system during and after interaction with the field as
\begin{flalign}
&\hat{\rho}_{\rm MATTER}(t) = {\rm Tr}_{\rm LIGHT}\left[\mathcal{P}(t-t_0)\hat{\rho}_{\rm TOT}(t_0)\right]\label{eq:basic}
\end{flalign}
where $\mathcal{P}$ is an operator determined by the internal structure of the system and its coupling to both the incident field and amplification processes, and $\hat{\rho}_{\rm TOT}(t_0)=\hat{\rho}_{\rm LIGHT}(t_0)\otimes\hat{\rho}_{\rm MATTER}(t_0)$ represents the initial density operator for the combined matter and field quantum state. ${\rm Tr}_{\rm LIGHT}$ represents a partial trace over the field degrees of freedom. This allows us to represent measurement outcomes $\Pi(t)$ as 
\begin{flalign}
&\Pi(t) = {\rm Tr}_{\rm LIGHT}\left[\mathcal{K}(t-t_0)\hat{\rho}_{\rm TOT}(t_0)\right]\label{eq:basic2}
\end{flalign}
which can be used to calculate the performance metrics.

For a monochromatic, single mode wavepacket, we write the incoming light field density matrix as 
\begin{flalign}
\hat{\rho}_{\rm LIGHT}(t) = \sum_{N,M} c_{N,M}(t)\ket{N}\bra{M},
\end{flalign} 
where $\ket{N}$ is a Fock state of the mode with occupation $N$.  The reduced density matrix for the matter degrees of freedom is
\begin{flalign}
&\hat{\rho}_{\rm MATTER}(t) = \sum_{N,M} c_{N,M}(t_0)\hat{\varrho}^{N,M}(t)
\end{flalign}
where  $\hat{\varrho}^{N,M}(t)$ are auxiliary matrices defined by the hierarchy
\begin{flalign}
\dot{\hat{\varrho}}^{N,M}(t)&=-i[\hat{H},\hat{\varrho}^{N,M}(t)]+\sum^{\rm BATHS}_i\mathcal{D}[\hat{Y}_i]\hat{\varrho}^{N,M}(t)\nonumber\\
&+\sqrt{N}E(t)\expn{-i\omega t}[\hat{\varrho}^{N-1,M},\hat{L}^\dagger_{i}]\nonumber\\
&\hspace{0cm}+\sqrt{M}E^*(t)\expn{i\omega t}[\hat{L}_{i},\hat{\varrho}^{N,M-1}]+\mathcal{D}[\hat{L}]\hat{\varrho}^{N,M}\nonumber\\
&+\sum^{\rm AMPS}_i\mathcal{D}[(2k_i)^{1/2}\hat{X}_i]\hat{\varrho}^{N,M}(t)\nonumber\\
&+\frac{dW_{i,t}}{dt}(2k_i)^{1/2}\left(\hat{X}_i\hat{\varrho}^{N,M}+\hat{\varrho}^{N,M}\hat{X}_i^\dagger-2\langle \hat{X}_i\rangle \hat{\varrho}^{N,M}\right)
\label{eq:full_me}
\end{flalign}
with ``BATHS'' and ``AMPS'' referring to sums over the number of baths and amplification channels respectively. The first term in this equation represents evolution of the matter/detector degrees of freedom according to its free Hamiltonian $\hat{H}$ and the second term represents non-unitary Lindblad evolution of these degrees of freedom as a result of coupling to uncontrolled bath modes (e.g., phonon modes in a material). The operators $\hat{Y}_i$ are the detector degrees of freedom that are coupled to the bath modes, and $\mathcal{D}[A]\rho \equiv A\rho A^\dagger - \frac{1}{2}A^\dagger A \rho - \frac{1}{2}\rho A^\dagger A$ is the generic Lindblad superoperator. The third through fifth terms represent the coupling of the detector to the photon pulse through the operator $\hat{L}$, with the photon pulse described by temporal profile $E(t)$. Finally, the last two terms model amplification of certain degrees of freedom of the detector (those represented by operators $\hat{X}_i$), with the sixth term modeling amplification-induced backaction and the seventh modeling the stochastic evolution of the system as a result of conditioning of a measurement record that results from the amplification. $W_{i,t}$ represents Weiner processes that model the intrinsic quantum noise (shot noise) associated with
amplification and measurement. This equation represents the evolution of the matter subsystem conditioned on
corresponding measurement records (such as electronic currents)
\begin{flalign}
I_i(t)=\frac{1}{t_m}\int^t_{t-t_m}dt'\left(\langle \hat{X}_i\rangle(t')+\frac{1}{(8k)^{1/2}}\frac{dW_{i,t'}}{dt'}\right)\label{eq:record}
\end{flalign}
where $k$ corresponds to the ``rate'' at which the measurement $\hat{X}_i$ is performed and $t_m$ is an integration time. One can also obtain the average evolution of the matter subsystem by averaging
over the measurement records.

By vectorizing the density matrix and recasting the superoperators as matrix operators \cite{Young:2018b}, we find that in Equation \eqref{eq:basic}, for the average dynamics,
\begin{flalign}
\mathcal{P}(t)=1+&\int_{0}^t\bar{\mathcal{G}}(t-\tau)
\hspace{0cm}\left(\bar{\mathcal{L}}^+dB_{\tau}+\bar{\mathcal{L}}^-dB^\dagger_{\tau}\right)\mathcal{P}(\tau)d\tau
\label{eq:P}
\end{flalign}
where the quantum noise operator field is defined as $dB_{\tau}\ket{N}=NE(\tau)\ket{N-1}$, and
$\bar{\mathcal{L}}$ is the field coupling and barred quantities represent vectorized forms (e.g., operators, decorated originally with hats, acquire bars when treated as vectors).

A key feature of Equation (\ref{eq:P}) is the Green's function $\bar{\mathcal{G}}$ which describes the internal properties of the matter subsystem. This function is obtained from the detailed Hamiltonian that describes the
system under consideration, and is what allows us to identify the limitations to existing photodetector designs
and propose new ones.

For an amplifier described by $\hat{X}_i=\chi_{i}\hat{x}_i$, where $\hat{x}_i$ is a projection operator, the probability that the amplifier registers an event (\ie~the measurement outcome) can be written as
\begin{flalign}
\Pi_{i}(t) =& \bar{x}_{i}\dg \bar{\mathcal{G}}(t-t_{\rm MIN})\bar{x}_{i}\nn\\
&\times\left(\bar{x}_{i}\dg\bar{\rho}(\tau)- \bar{x}_{i}\dg \bar{\mathcal{A}}\bar{x}_{i}\int_{t_0}^{t-t_{\rm MIN}} d\tau~ \bar{x}_{i}\dg\bar{\rho}(\tau)\right)\label{eq:hitprob}
\end{flalign}
where $t_{\rm MIN}$ is the minimum time that the state $i$ must be occupied for a detection event to be registered. The above expression is modified when one desires the joint probability that multiple channels register events.  For example, for the joint probability $\Pi_{ijk}(t)$ that three channels $i,j,k$ have registered events, the above is modified by $\bar{x}_i\rightarrow\bar{x}_i\otimes\bar{x}_{j}\otimes\bar{x}_k$.  More complicated expressions may be easily, if tediously, derived for amplification operators comprising multiple projectors.

In general, the outcome we are interested in will depend on the status of multiple detection channels and will thus be a function of all channel outputs.  For example, we may have a set of detection channels like the above, each monitoring the system for entry into a state where a photon has been absorbed by the detector.  Then, the outcome of a single photon being detected is simply the sum over all the channel outputs $P_1(M,t)=\sum_i\Pi_i(t)|_M$, where we introduce the notation $P_N(M,t)$  for the probability that $N$ photons have been detected at time $t$ given an input signal with $M$ photons.  Note that this quantity depends on the field profile $E(t)$, but we omit it in the notation for brevity. Multiple photons require summation over the joint probabilities described above; \eg the outcome of three photons detected is $P_3(M,t)=\sum_{i\ne j\ne k}\Pi_{ijk}(t)|_M$.   From this we can determine a number of key metrics.   

{\it Efficiency}. Efficiency is  the probability of having detected all photons once the light pulse has completely passed, which we will denote as $P_N=P_N(N,\infty)$.
This efficiency for $N>1$ is the metric we shall identify with photon number resolving ability.  We note that in principle we can more completely characterize photon resolution performance by considering $P_{N-1}(N,\infty)$, $P_{N-2}(N,\infty)$, etc., but for simplicity we will not do so here. 

{\it Jitter}. Since $\dot{P}_N(M,t)$ gives a distribution of detection times, the total jitter for the $N$th photon can be computed as
\begin{flalign}
\sigma&=\sqrt{\int_{t_0}^\infty dt\, t^2\frac{\dot{P}_N(N,t)}{P_N}-\left(\int_{t_0}^\infty dt\, t\frac{\dot{P}_N(N,t)}{P_N}\right)^2}.
\end{flalign}
For convenience we define $\sigma_{{\rm SYS}}$ as the total jitter minus the temporal width of the pulse
\begin{flalign}
\sigma&=\sqrt{\left(\sigma_0\right)^2+\left(\sigma_{{\rm SYS}}\right)^2}
\end{flalign}
with
\begin{flalign}
\sigma_0&=\sqrt{\int_{t_0}^\infty dt\, t^2\dot{f}(t)-\left(\int_{t_0}^\infty dt\, t\dot{f}(t)\right)^2}\\
f(t)&=\int^{t}_{t_0} d\tau\abs{E(\tau)}^2.
\end{flalign}

{\it Dark count rate}. The dark count rate intrinsic to the detector is given by 
\begin{flalign}
r_{i}=\frac{\Pi_i(0,t_m+t_0)}{t_m}+\frac{0.5}{t_m}{\rm erfc}\left(2\sqrt{kt_m}\Delta I_{{\rm HIT},i}\right)
\end{flalign}
where $\Pi_{i}(0,t_m+t_0)$ is the probability of obtaining a hit in time $t_m$ due to internal detector dynamics when no photons are present in the field, the second term is due to shot noise, and $\Delta I_{{\rm HIT},i}$ is the difference in signal between hit and non-hit states of the detector.  We note that the shot noise included in the above expression is fundamental to the measurement process and separate from additional noise introduced by downstream transmission and processing of the classical signal.  It is thus convenient to define a maximum signal-to-noise ratio determined by this intrinsic noise, SNR$_{0,i}=\sqrt{8kt_m}\Delta I_{{\rm HIT},i}$.

{\it Count rate}. For a detector to continuously register incident photons, the detector state must reset after an event is recorded.  If the detector passively resets according to some internal mechanism wherein it returns to the ground state, we can take the rate of this mechanism as determining the maximum count rate. As discussed below, this mechanism is naturally included in $\bar{\mathcal{G}}$.

{\it Bandwidth}. The bandwidth is defined as the photon energy range over which the detector maintains its efficiency above a certain pre-determined value.

\section{The need to design for high performance PNR}

Previously \cite{Young:2018b} it was demonstrated that the above formalism can be used
to identify a basic structure for a detector element that achieves perfect efficiency, no dark counts,
and minimal jitter for {\it single} photon detection.
As illustrated in Figure~2a, the system consists of a two-level subsystem with photon absorption rate $\gamma$, a transduction subsystem that incoherently transfers the excitation into a shelving state at rate $\Gamma$, and the continuous measurement process with strength $\chi$.  In terms of Equation \eqref{eq:full_me} these processes are represented by the operators 
\begin{eqnarray}
\hat{H}=\omega_0\ket{1}\bra{1}, \hat{L}=\gamma\ket{0}\bra{1}, \hat{Y}^\Gamma=\Gamma\ket{C}\bra{1}, \nonumber \\
\hat{Y}^\Delta=\Delta\ket{0}\bra{C}, \hat{X}=\chi\ket{C}\bra{C}.
\end{eqnarray} 
Ideal detection occurs when both $\gamma = \Gamma \gg 1/\sigma_0$ and SNR$_0=\sqrt{8kt_m}\chi\gg 1$. 

\begin{figure*}
	\includegraphics[width=0.7\textwidth]{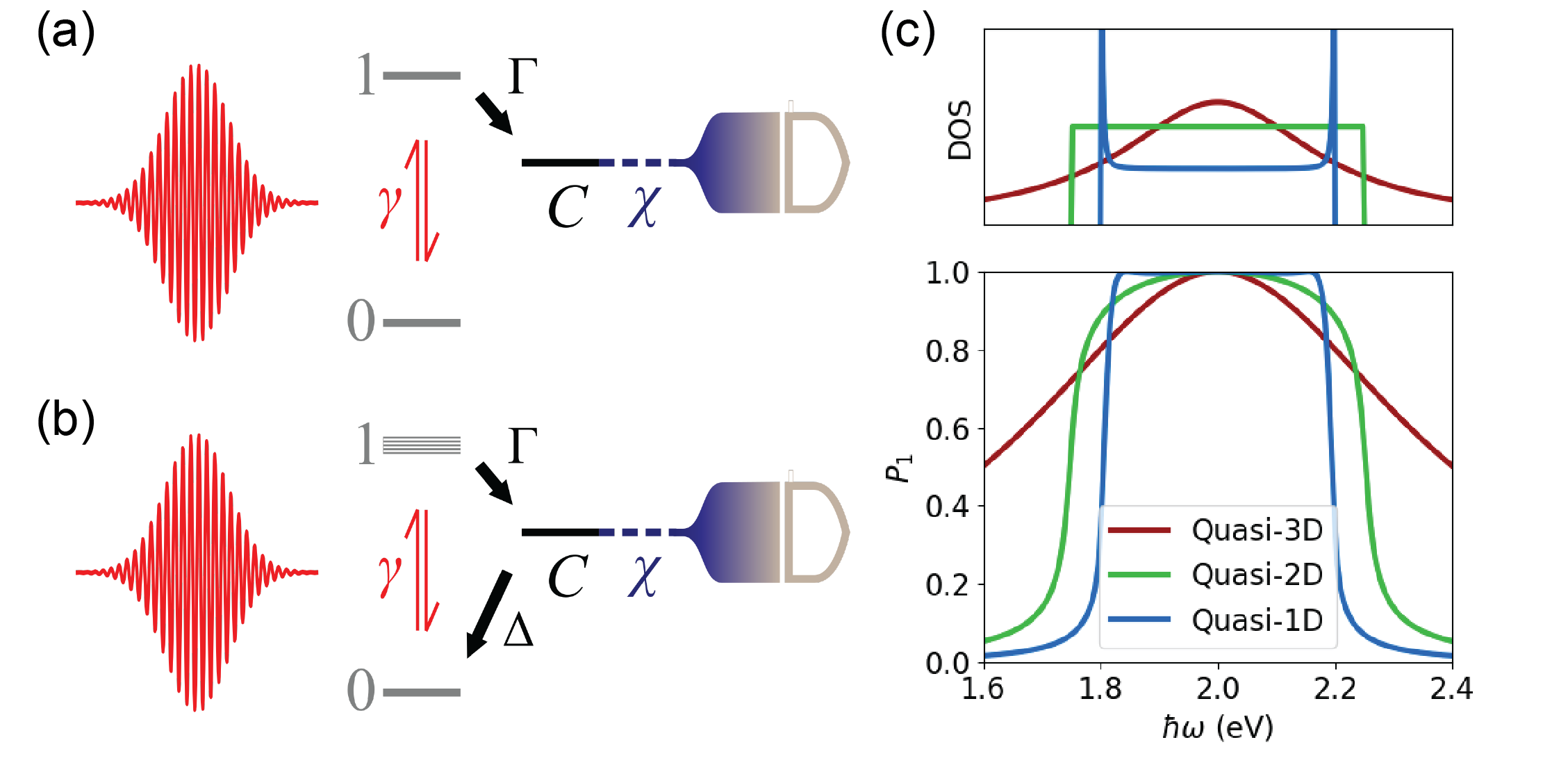}
	\centering
	\caption{ Photodetector element structure and performance. a) Element structure that gives
	high efficiency, low jitter, and low dark counts for {\it single} photon detection when $\gamma=\Gamma$. b) Element structure that adds band absorption in 1 and decay of the $C$ state to achieve high count rate and high bandwidth. c) Electronic densities of states (top) for three different dimensionalities and the associated detector efficiency (bottom) assuming $\Delta=0$.} 
	\label{fig:sys2}.
\end{figure*}

While this detector element achieves high performance, it still lacks in two of the metrics: count rate and
bandwidth. Indeed, the structure in Figure~\ref{fig:sys2}a does not contain an intrinsic mechanism to reset the detector
and thus achieve a high count rate. In addition, the bandwidth is low due to optical absorption by two narrow states. So at least, the individual detector elements need to be designed to
additionally perform well in these two metrics.

Furthermore, the whole photodetector needs to be designed to achieve PNR. The simplest way to achieve PNR is to multiplex a number of these detection elements. Unfortunately, 
whether such elements interact separately \cite{Jonsson:2018} or collectively with the field \cite{Young:2018b},
a large number of elements is necessary to achieve number resolution and high efficiency up to the $N$th photon. In the case of an array of $n$ collectively interacting elements, while ideal \emph{single} photon efficiency may be preserved by letting $\sqrt{n}\gamma=\Gamma$, once one element has
absorbed a photon, subsequent photons interact with a detector containing one fewer elements, detuning the array and reducing efficiency. While this effect can be suppressed by making $n$ larger, the required $n$ increases rapidly with desired photon resolution capability. For example,
number resolution of 50 photons with a detection efficiency of 99\% for the 50th photon requires more than 1000 elements even with perfect single photon detectors.  Since each element includes an amplifier, this has a direct impact on device complexity and noise. The challenge therefore is to improve on this scaling while also achieving high performance
in all the other metrics. We now present our step-by-step approach to design such a system.

\section{Designing for bandwidth}
A useful way to increase bandwidth is to move beyond two-level absorption systems and consider
systems with a distribution of electronic states, such as molecules, quantum dots, two-dimensional
materials, or three-dimensional solids, leading to the modified structure of Figure~\ref{fig:sys2}b. 
In terms of Equation \eqref{eq:full_me} this corresponds to 
\begin{eqnarray}
\hat{H}=\sum_{l}\omega_l\ket{1_l}\bra{1_l}, \hat{L}=\sum_{l}\gamma_l\ket{0}\bra{1_{l}}, \hat{Y}^\Gamma_{l}=\Gamma_{l}\ket{C}\bra{1_{l}}\nn\\
\hat{Y}^\Delta=\Delta\ket{0}\bra{C}, \hat{X}=\chi\ket{C}\bra{C}
\label{eq:band_ops}
\end{eqnarray} 
where $l$ indexes the states in band 1, $\gamma_l$ are the associated optical couplings, and $\Gamma_l$ are the incoherent decays. The probability of detection in such a system depends on the relationship between the parameters $\gamma$, $\Gamma$ and the shape of band $1$.  In the continuum limit when the density of states  (DOS) of the band is described by a Lorentzian with FWHM $\zeta^2$ and the $\gamma_l$ and $\Gamma_l$ are the same for all states in 1, it is possible to compute the probability of detection analytically as 
\begin{flalign}
P_1(1)=\frac{4n_b\gamma^2(\Gamma^2+\zeta^2)}{(n_b\gamma^2+\Gamma^2+\zeta^2)^2}\frac{1}{1+\frac{4\delta\omega^2}{(n_b\gamma^2+\Gamma^2+\zeta^2)^2}} \label{eq:band_ideal}
\end{flalign}
where $\delta\omega$ is the detuning from the band center frequency, $n_b$ is the number of states in $1$, and we have assumed $\Delta=0$ for now~\cite{Young:2018b}.  It is clear that when $n_b\gamma^2=\Gamma^2+\zeta^2$ the probability at the band center is unity. This result indicates that it is possible to achieve high
efficiency while introducing some bandwidth; however, the efficiency decreases steadily away from the
center frequency, thus limiting the bandwidth. To illustrate this effect, we show in Figure~\ref{fig:sys2}c a numerical result for the case of the Lorentzian distribution, representing three-dimensional solids. 

It turns out that the shape of the DOS has an important effect on the bandwidth. Indeed,
Figure~\ref{fig:sys2}c also shows results for the case of two-dimensional materials where the DOS is
constant over some energy range, which already shows improvements over the three-dimensional case, \ie~
high detection efficiency over a wider energy range.
But the real impact is seen when going to quasi-one-dimensional systems, where the DOS
contains divergences at the band edges (e.g.~van Hove singularities). There, a perfect detection efficiency
is maintained over almost the whole DOS bandwidth.

It is important to note that this result is not restricted to constant $\gamma_l$ and $\Gamma_l$.  In fact, in the Supporting Information we show that unit efficiency at least one frequency is possible as long as the following conditions are satisfied: 1) relaxation processes do not couple optically active processes to one another, 2) relaxation processes do not couple dark states back to the optically active manifold, and 3) the spectral distribution of optical intensity is symmetric.  Even when these are violated and lead
to losses, the impact on performance may be minimal, as we will show below.

This result has implications for the physical realization of these detector elements: the absorption
process should occur in quasi-one-dimensional materials such as extended molecules, carbon nanotubes,
or atomic chains.

\section{Count rate}
In order to achieve high count rate, we introduce an internal relaxation mechanism from the $C$ state to the ground state (Figure~2b), allowing the detector to reset and become available to detect another photon.  However, this also introduces the possibility for the detector to reset before the minimum measurement interval $t_{\rm MIN}$ has elapsed, resulting in no detection and efficiency loss.  In general for the above system, using Equation \eqref{eq:hitprob} we can write 
\begin{flalign}
P_1(1)=\expn{-\Delta t_{\rm MIN}}\frac{4n_b\gamma^2(\Gamma^2+\zeta^2)}{(n_b\gamma^2+\Gamma^2+\zeta^2)^2}\frac{1}{1+\frac{4\delta\omega^2}{(n_b\gamma^2+\Gamma^2+\zeta^2)^2}}.
\end{flalign}
Furthermore, for a given count rate $r_{\rm C}$, the probability that the detector has failed to reset in time to detect the next incoming photon is given by $\expn{-\Delta/r_{\rm C}}$, further reducing the efficiency.  Overall the efficiency is then reduced by a factor of ${\rm Eff}_{\rm LOSS}=1-\expn{-\Delta t_{\rm MIN}}(1-\expn{-\Delta/r_{\rm C}})$.
Thus, the maximum count rate attainable for a given ${\rm Eff}_{\rm LOSS}$ is
\begin{flalign}
r_{\rm C}&=-\frac{\Delta}{\ln \abs{1-(1-{\rm Eff}_{\rm LOSS})\expn{\Delta t_{\rm MIN}}}}\nn\\
 &\approx-\frac{\Delta}{\ln\abs{{\rm Eff}_{\rm LOSS}+\Delta t_{\rm MIN}}}\label{eq:single_rc}
\end{flalign}
where the approximate expression is valid when ${\rm Eff}_{\rm LOSS}\ll 1$. 

We see that for this detector, even allowing $t_{\rm MIN}\rightarrow 0$,  there is an unavoidable tradeoff between ensuring high efficiency and achieving high count rate.

\section{Designing for PNR}

Having designed improved high performance detector {\it elements} for {\it single} photon detection, we now consider arrays of such elements for PNR. As discussed
above, simple multiplexing of single photon detection elements requires a large number of elements to achieve PNR
and high performance. Since each element entails an amplification process, increasing the number of elements increases both noise and device complexity.  

The central issue is the unavailability of a detection element once it has absorbed a photon, and
the impact on the collective absorption strength. Introducing a decay of the $C$ state to the ground state, clearing the element, does not solve this
problem on its own because the $C$ state needs to be populated for a minimum amount of time for measurement to yield a signal above the noise.
A solution to this problem is to couple the decay of the $C$ state to the excitation of an optically inactive secondary system via an energy transfer process. This achieves two goals: it
returns the system to its absorption ground state, maintaining a consistent detector coupling to the field as photons are absorbed, and transfers the excitation to a subsystem that
can be monitored for an arbitrary long period of time and need only contain enough amplifiers for the desired photon resolution, preserving the signal to noise ratio.

\begin{figure*}
	\includegraphics[width=0.7\textwidth]{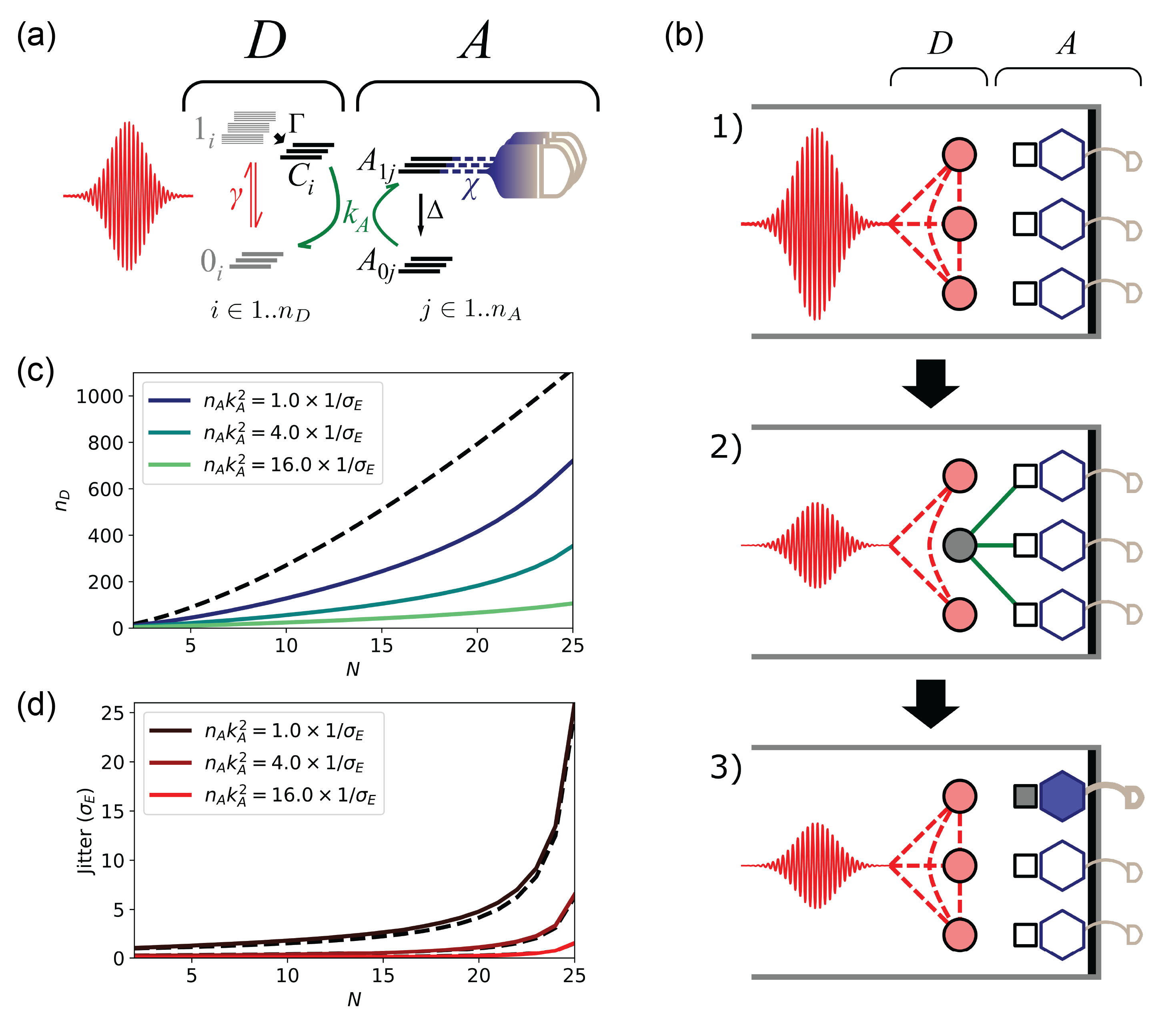}
	\centering
	\caption{  Structure for a high performance photon number resolving detector and performance. a)-b) A pulse containing $N$ photons is absorbed at rate $\gamma$ by subsystem $D$ which contains $n_D$ absorbing elements. This excites $N$ elements in $D$ from their ground state 0 to their excited band 1. Each excited element transfers the excitation at rate $\Gamma$ to its shelving state $C$. The population in $C$ is then transferred to subsystem $A$ that contains $n_A$ elements. This excites $N$ of these elements
from their ground state 0 to their excited state 1. Each excited state element in $A$ is monitored by an amplifier at rate $\chi$, and naturally decays at rate $\Delta$. c) The minimum number of idealized absorbing elements $n_D$ needed to achieve 99.9\% efficiency for a given number of photons assuming $n_A=25$ and $\Delta=0$. The dashed line is the case of a simple array of uncoupled elements. d) Jitter as a function of photon number for different energy transfer rates. The dashed lines are fits as described in the text. 
	  \label{fig:sys_K}}
\end{figure*}

Taken together, these considerations suggest the detector architecture represented in Figure~\ref{fig:sys_K}a.
It is characterized by two subsystems, labeled $D$ and $A$. In general, the overall system may contain $n_D$ copies of $D$ and $n_A$ copies of $A$, each described by its own Hilbert space.  The $D$ systems are essentially unamplified versions of the system in Figure~\ref{fig:sys2}b, so that their mathematical description (Equation \eqref{eq:band_ops}) becomes
\begin{eqnarray}
\hat{H}_i=\sum_{l}\omega_l\ket{1_{il}}\bra{1_{il}}, \hat{L}=\sum_{il}\gamma_l\ket{0_i}\bra{1_{il}}, \nn \\
\hat{Y}^\Gamma_{il}=\Gamma_{l}\ket{C_i}\bra{1_{il}}
\end{eqnarray} 
where $i$ indexes the $D$ subsystems.  We note that $\hat{L}$ couples light to each of these subsystems, resulting in coherent excitation of the combined system of elements and a new condition for optimal efficiency obtained by taking $n_b\gamma^2\rightarrow n_Dn_b\gamma^2$ in Equation \eqref{eq:band_ideal}.
  
The $A$ subsystems are characterized by a single ground state and excited state and are amplified by a mechanism that projects onto the excited state.  An $A$ system will naturally experience a decay from the excited to ground state at rate $\Delta$, serving as a reset mechanism.  The two types of subsystem are coupled via a one way exchange of energy that returns a $D$ system to the ground state from the $C$ state while promoting an $A$ system to the excited state.  Thus
\begin{flalign}
\hat{Y}^\Delta_i&=\Delta\ket{A_{0j}}\bra{A_{1j}} \nn \\
\hat{Y}^{k_A}_{ij}&=k_A\ket{0_i}\bra{C_i}\otimes\ket{A_{1j}}\bra{A_{0j}}
\end{flalign} 
where $A_{0j}$ and $A_{1j}$ are the ground and excited states of the $j$th $A$ subsystem and $k_A$ is the energy transfer coupling.

Thus, multiphoton detection proceeds as depicted in Figure~\ref{fig:sys_K}a.  First, an incoming pulse containing $N$ photons interacts coherently with the collection of $D$ subsystems.  Next, an incoherent decay occurs, leaving one of the $D$ systems in its $C$ state, while the other $n_D-1$ systems continue interacting with the remaining $N-1$ photons.  Photons may continue to be absorbed, further reducing the number of $D$ systems interacting with the field.   However, concurrently the occupied $D$ systems may transfer their energy to one of the $A$ systems (which undergo amplification) restoring the availability of the $D$ system.  If the rate at which energy is transferred and the $D$ systems restored is rapid enough, the number of occupied $D$ systems is minimized and the number participating in absorption is maintained near $n_D$. 

We will now show quantitatively that this abstract structure attains high performance and discuss the parameter regimes that optimize the performance.  We will then analyze more specific realizations and show that
molecular systems coupled with nanoscale device elements can achieve high performance PNR.

\subsection{Energy Transfer}

Intuitively, it would seem that for sufficiently high rates of energy transfer, the described system can efficiently collect photons while maintaining itself in an optimal state for interaction with the field. Additionally, the number of required amplifiers is restricted to the number of acceptor systems $n_A$, while $n_D$ may be made arbitrarily large to both increase multiphoton efficiency as well as tune the detector to the optimal regime defined by Equation \eqref{eq:band_ideal}.  To see the requirements for $n_D$ and $k_A$, we plot in Figure~\ref{fig:sys_K}b the required number of idealized absorbing elements to fully resolve a multiphoton packet 99.9\% of the time as a function of the number of photons in the wavepacket for varying overall energy transfer rates $n_Ak_A^2$ and $n_A=25$. The energy transfer rates are given in terms of the inverse of the photon pulse width; \ie~the rate at which photon amplitude enters the system.  We see that faster energy transfer does indeed reduce the need for large $n_D$.  

The jitter also depends critically on both $k_A$ and $n_A$.  In Figure~\ref{fig:sys_K}c the jitter is plotted as a function of incoming photon number.  We see that for small numbers of photons the jitter introduced by the energy transfer process is $\sim 1/(n_Ak_A)$.  However, it rises dramatically as the photon number approaches $n_A$.  This can be understood as a saturation effect; as the number of photons absorbed increases the number of $A$ systems available for energy transfer decreases, reducing the overall rate and increasing the variance in the time needed for energy transfer to occur.  Indeed the data is
well described by the expression
\begin{flalign}
\sigma = \frac{ \sigma_0 n_A}{n_A k_A^{2} (n_A -N +1)}
\end{flalign}
indicating that minimization of jitter requires operation far from saturation. For example, if a maximum of $N$ photons need to be detected, then operating at $n_A=2N$ gives only about twice the jitter of single photon absorption. We will use this condition further below.

\subsection{Amplification and count rate}
The introduction of the $A$ subsystems has the added effect of serving as a buffer, allowing much higher count rates.  Successful detection is no longer contingent upon any particular $A$ system having reset following prior detection; photons can continue to be detected efficiently as long as there are available $A$ systems to accommodate the additional photons without exceeding the required jitter. Specifically,  the maximum count rate is equal to the rate at which $A$ systems are reset, which at the desired number of $A$ systems $n_A = 2N$ gives $r_{\rm C}=N\Delta$ since each $A$ system resets at a rate $\Delta$.  Additionally, we now have ${\rm Eff}_{\rm LOSS}=1-\expn{-N\Delta t_{\rm MIN}}$, which is the efficiency loss for detecting $N$ photons due to the possibility that any excited acceptor site decays.  As a result,
\begin{flalign}
r_{\rm C}\approx N \frac{{\rm Eff}_{\rm LOSS}}{t_{\rm MIN}}.
\end{flalign}
The count rate can now be made arbitrarily high as long as $t_{\rm MIN}$ can be reduced,
a considerably more favorable scenario than for Equation \ref{eq:single_rc}.  
However, there are practical limits to $t_{\rm MIN}$ that can arise as a result of dark count requirements.
Since each $A$ system is monitored by an amplifier, letting $t_m=t_{\rm MIN}$ we obtain
\begin{flalign}
r_{\rm DC}&=\sum_i^{n_A}r_i=\frac{n_A}{2t_{\rm MIN}}{\rm erfc}\left(2\sqrt{kt_{\rm MIN}}\chi\right) \nn \\
&=\frac{N}{t_{\rm MIN}}{\rm erfc}\left(\frac{{\rm SNR}_0}{\sqrt{2}}\right)
\end{flalign}
where, for the amplification under consideration, $\Delta I_{{\rm HIT}}=\chi$ according to Equation \eqref{eq:record}, and we have assumed that $\frac{\Pi_i(0,t_{\rm MIN}+t_0)}{t_{\rm MIN}}\approx 0$; \ie~that random excitations of the $A$ subsystem occur at a negligible rate. It is evident that shorter measurement times both increase the required SNR$_0$ for a given dark count rate and directly decrease SNR$_0$. 
As a result, increased count rates come at the expense of increased dark count rates.  Specifically, 
\begin{flalign}
\frac{r_{\rm DC}}{r_{\rm C}}&\approx\frac{1}{{\rm Eff}_{\rm LOSS}}{\rm erfc}\left(\frac{{\rm SNR}_0}{\sqrt{2}}\right)
\label{eq:dcmc}.
\end{flalign}
We note that this tradeoff is ultimately constrained by the overall magnitude of amplification $\sqrt{2k}\chi$, which is determined by the physical mechanism; stronger amplification in this case allows for both fewer dark counts and higher maximum count rates.

\section{Physical Realization and Parameters}
\begin{figure*}
	\includegraphics[width=0.8\textwidth]{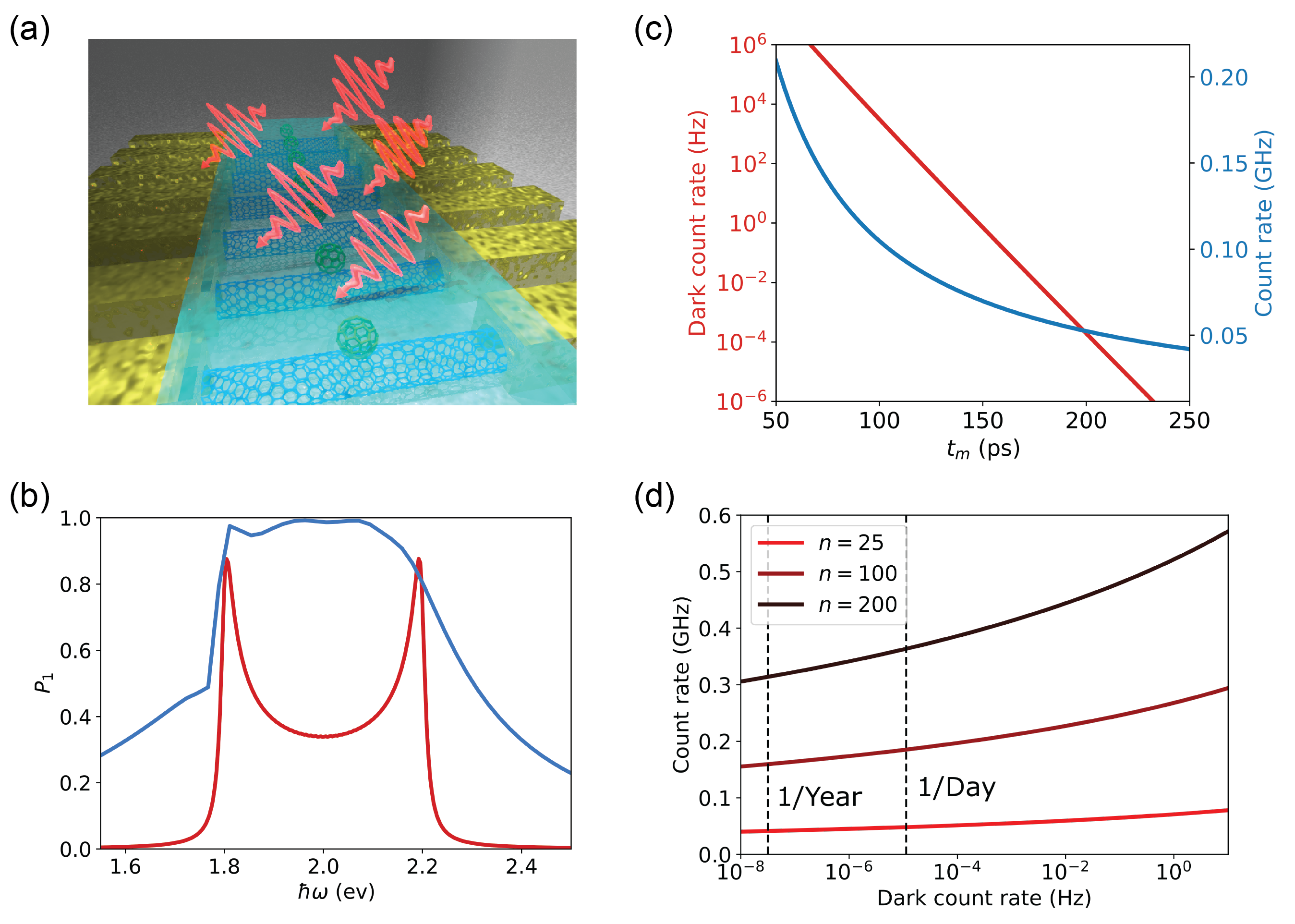}
	\centering
	\caption{ Physical realization of the new detector design. a) Photons are
	absorbed by a molecular thin film (cyan) creating excitons. The excitons are transfered to acceptor
	molecules or quantum dots (e.g.~C$_{60}$) that functionalize quantum transport channels (e.g.~a carbon
	nanotubes between metallic leads). The excitons are dissociated, leaving trapped charges that modulate the current in the channel. The detector sits in a single mode waveguide. b) The density of optical activity for transitions to e-h pair states (red) of a model $D$ system and the resulting efficiency upper bound (dark blue).  An exciton binding energy of 220meV, a phonon energy of 40meV, an oscillator strength of 0.25, and $n_d=2.5\times 10^5$ were assumed. c) Trade-offs between dark count rate, integration time, and count rate for 99\% efficiency of detecting 12 photons with $n_A=25$ and assuming the $D$ system is performing ideally.  d) Count Rate vs. Dark Count Rate for varying $n_A$, assuming 99\% efficiency for 12 photons assuming ideal absorption.}
	\label{fig:detector}
\end{figure*}

The above requirements lead to candidates for physical realizations.  The need for a quasi-one-dimensional absorbing element to experience incoherent decay to a relatively long-lived state, as well as a mechanism for energy transfer, suggests that molecules, quantum dots or similar systems hosting excitons are suitable absorbers. When physically collocated with appropriate secondary molecules, fluorescence resonance energy transfer can occur with high efficiency and high rate between the absorber subsystem ($D$) and the acceptor subsystem ($A$). The resulting excitation of the acceptor molecules can be monitored by
putting them in close proximitiy to high performance quasi-one-dimensional nanoscale electronic transport
channels \cite{bergemann, Weng_2015}. This leads to the design in Figure~\ref{fig:detector}a. 

First we consider the implications for the $D$ systems. The parameters that describe this system are the 
incoherent transfer rate $\Gamma$, the excitation bandwidth $\zeta$, the number of absorbing molecules $n_D$, and the effective absorption rate $\gamma_{\rm Eff}$. Here $\gamma_{\rm Eff}$ is the optical absorption rate in
the particular optical environment given $n_D$ absorbing molecules of DOS bandwidth $\zeta$. For simplicity, we will consider a Lorentzian DOS since analytical results can be readily obtained in this case.

The condition for high performance is
\begin{flalign}
\gamma^2_{\rm Eff}=\Gamma^2 + \zeta^2 \approx \zeta^2
\label{condition}
\end{flalign}
where the last approximation comes from the fact that typical phonon-mediated incoherent processes
occur on time scales of 100 fs, compared to DOS bandwidths of 1fs. 

The total spontaneous emission rate in a single-mode waveguide is related to the free space spontaneous emission rate $\gamma^2_{\rm Free}$ for a single element as (Supporting Information)
\begin{flalign}
\gamma^2_{\rm Eff}\approx \frac{3\lambda^2}{4\pi A_{\rm WG}}n_D\gamma^2_{\rm Free}
\end{flalign}
where $A_{\rm WG}$ is the cross sectional area of the waveguide and $\lambda$ is the wavelength of the mode in question. We note that the exact relation depends on the waveguide details. The free space spontaneous emission rate can be related to the molecular absorption cross section through
\begin{flalign}
\gamma^2_{\rm Free}\approx\frac{4}{\lambda^2}\frac{\sigma(\omega)}{\rho(\omega)}
\end{flalign}
where $\rho ( \omega )$ is the DOS~\cite{Hillborn_1982}. For the Lorentzian case, $\rho ( \omega )=(\zeta^2 / (2\pi))/[\omega^2+(\zeta^2/2)^2]$ and using the condition in Equation (\ref{condition}) we obtain at the resonant frequency

\begin{flalign}
n_D\approx\frac{ 2A_{\rm WG}}{3\sigma(\omega_0)}.
\end{flalign}

Given that optical cross section areas for molecules are typically on the order of 1-10$\AA^2$  we obtain $n_D\approx 10^4-10^6$. Alternatively for a thin film of thickness $h$ that spans the waveguide, we can use the optical absorption coefficient $\alpha$ to obtain $h\approx\frac{2}{3\alpha(\omega_0)}$.
For typical absorption coefficients in the range $1-5\times 10^7$m$^{-1}$, this suggests that thin films need only be a few tens of nanometers in thickness to achieve high performance.

To assess the performance potential of excitonic systems more specifically, we constructed a representation of an 1D excitonic systems that experiences phonon mediated relaxation and evaluated its single photon detection performance.  We assume a 32-site, two-band tightbinding model with Coulomb interactions approximated by the Ohno potential and coupling to local phonons.  This construction and its properties are detailed in the Supporting Information. With parameters chosen to represent realistic systems, we calculated the single photon detection efficiency for the architecture of Figure~2b with no reset. The result in Figure~4b confirms that high efficiency is possible despite imperfectly satisfying the conditions for ideal detection. Since the PNR detector is constructed from efficient sigle photon detector elements, this result suggests that excitonic systems are appropriate elements for system $D$.

We now discuss the energy transfer rate. As mentioned above, the number of $A$ elements $n_A=2N$ is chosen to minimize jitter, with values of $k_A^2n_A$ comparable to the pulse width giving low jitter. Therefore we have the condition $k_A^2 = 1/(2N\sigma_0)$. Thus, as a consequence of minimizing jitter, the needed transfer rate decreases with the maximum number of photons to be resolved. One mechanism for energy transfer is F\"{o}rster resonant energy transfer (FRET), with time scales on the order of picoseconds \cite{Wang_2019} and efficiencies that can attain near 100\% \cite{Kreisig_2013}, which can be further tuned with the optical
environment \cite{Cortes_2018}. This implies that the configuration of Figure~\ref{fig:detector} would be able to resolve photon number as long as the pulse is longer than a picosecond. (Losses due to this process will
impact the overall performance and will be considered in more detail in the future.)

Once the exciton is transferred to the A susbsystem, it needs to be monitored by a measurement process. Since directly measuring excitons is difficult, it is useful to dissociate the exciton into electrons and holes using built-in electric fields. One approach is to use the band aligment at the interface between molecules and quantum transport channels, which leaves a long-lived hole or electron in proximity to the channel. Recent experiments and modeling have indicated that this mechanism
can be used to induce large modulations of carbon nanotube electrical conductance \cite{bergemann}. In that scenario, SNR$_0$ is given by
\begin{flalign}
{\rm SNR}_0(t_m)=f\sqrt{\frac{It_m}{2e}}
\end{flalign}
where the change in current $\Delta I = fI$ and we have assumed that the detector can be operated at high enough frequency
for shot noise to be dominant.

For example, carbon nanotubes can sustain currents as high as 10$\mu$A \cite{Franklin_2012}. Assuming a modest 10\% change in current upon exciton dissociation, we obtain ${\rm SNR}_0(t_m) \approx 10^{6}\sqrt{t_m}$.
Inserting this into our equations for dark count rate and max count rate and assuming an acceptable efficiency loss due to amplification of 1\%, we find the performance possibilities for $N=12$ in Figure~\ref{fig:detector}c.  We note in particular that a count rate of 50MHz can be achieved with a dark count rate of 1/day, provided that there are no additional sources of noise.
The situation can be improved by increasing $n_A$.  While this requires higher $t_{\rm MIN}$ to obtain the same overall dark count rate, Equation \eqref{eq:dcmc} shows that the accompanying improvement in SNR$_0$ nonetheless allows for increased maximum count rates for a given dark count rate.  We plot in Figure~\ref{fig:detector}d the maximum count rates vs~ dark count rates for different $n_A$ for less than 1\% efficiency loss.  Ultimately, it is advantageous to include as many $A$ subsystems as is practically feasible since it enables high performance in other metrics with minimal losses to efficiency, which, under the above circumstances is primarily determined by how close to ideal the $D$ subsystems are.

 \section{Conclusions}
 In summary, we employed a fundamental approach based on quantum master equations to identify the challenges
 in high performance photon numer resolving photodetectors. A number of obstructions arise when attempting to achieve PNR while simultaneously optimizing important metrics. 
  Using our approach we are
 able to understand the reasons for these obstructions and formulate designs that circumvent them. As a result, we designed a novel detector architecture
 based on coherently and collectively interacting absorbing elements, energy transfer, and a continuous monitoring process, that is able to achieve PNR as well as excellent performance in terms of efficiency, dark counts, bandwidth, and count rate.
 The needed physical properties of this architecture suggest that molecular and nanoscale systems are prime candidates
 to realize new generations of photodetectors.

\begin{acknowledgement}
Work supported by the DARPA DETECT program. Sandia National Laboratories is a multimission laboratory managed and operated by National Technology and Engineering Solutions of Sandia, LLC., a wholly owned subsidiary of Honeywell International, Inc., for the U.S. Department of Energy's National Nuclear Security Administration under contract DE-NA-0003525. The views expressed in the article do not necessarily represent the views of the U.S. Department of Energy or the United States Government.
\end{acknowledgement}

\bibliography{stoch_draft}



\section*{Supplementary material}

Supporting Information for

# Design of High-Performance Photon Number Resolving Photodetectors Based on Coherently Interacting Nanoscale Elements

Steve M. Young, Mohan Sarovar, François Léonard*

*Sandia National Laboratories, Livermore, CA, 94551, USA*

*Email: fleonar@sandia.gov

## 1. RELATIONSHIP BETWEEN $\gamma^2$ AND ABSORPTION CROSS SECTION

In order to estimate the number of donor elements $n_D$ that the absorbing subsystem must comprise, we wish to connect the parameters of the model for the absorbing system to observable quantities. We start by observing that the expression for the spontaneous emission rate $\gamma^2_{\text{WG}}$ in a rectangular waveguide [1] with width $L_x$ and height $L_y$ is

$$\gamma^2_{\text{WG}} = \frac{\lambda}{4\pi L_x L_y} \frac{1}{\sqrt{1 - \left(\frac{\lambda}{2L_x}\right)^2}} g^2$$

where $\lambda$ is the resonant wavelength and $g^2$ is the total optical coupling at the resonant frequency. The spontaneous emission rate in free space is

$$\gamma^2_{\text{FREE}} = \frac{2}{3\lambda} g^2$$

which leads to

$$\gamma^2_{\text{WG}} = \frac{3\lambda^2}{8\pi L_x L_y} \frac{1}{\sqrt{1 - \left(\frac{\lambda}{2L_x}\right)^2}} \gamma^2_{\text{FREE}}.$$

Choosing $L_x = \lambda/\sqrt{3}$ and $L_y = \lambda/\sqrt{12}$ we obtain

$$\gamma^2_{\text{WG}} = \frac{3\lambda^2}{4\pi A_{\text{WG}}} \gamma^2_{\text{Free}} = \frac{9}{2\pi} \gamma^2_{\text{Free}}.$$

## 2. CONDITIONS FOR IDEAL SINGLE PHOTON PERFORMANCE

In the main text we describe the conditions under which the system depicted in Fig. 2B can be tuned via $n_D$ to detect single photons with nearly perfect efficiency. Here we describe those conditions mathematically and show that these lead to a concise expression for efficiency that both manifestly allows for ideal detection and allows for straightforward determination of the optimal $n_D$. Assuming stable population in the amplified state ($\bar{x}_i^\dagger \bar{\mathcal{G}}(t) \bar{x}_i = 1$, which implies $\hat{L}\hat{X}_i = 0$), Eq. 6 of the main text becomes

$$\Pi_i(t) = \bar{x}_i^\dagger \bar{\rho}(t - t_{\text{MIN}}).$$

Additionally we assume that the amplification is efficient, *i.e.*, that the amplification channels span the space of the final stable states and their associated projectors are orthogonal, so that $\hat{Y}_j^\dagger \hat{Y}_k = \sum_i \hat{Y}_j^\dagger |x_i\rangle \langle x_i| \hat{Y}_k$. We can write at long times, assuming that $E(\omega)$ is centered on $\omega_0$ and sufficiently narrow compared to $\bar{\mathcal{G}}(\omega)$ around $\omega_0$ that it behaves

like a delta function,

$$\Pi_i(t) = \int_{t_0}^t d\tau \bar{x}_i^\dagger \bar{\mathcal{G}}(t-\tau)\left[\bar{\mathcal{L}}^+ E(\tau)\int_{t_0}^\tau d\tau' \bar{\mathcal{G}}(\tau-\tau')\bar{\mathcal{L}}^- E^*(\tau') + \bar{\mathcal{L}}^- E^*(\tau)\int_{t_0}^\tau d\tau' \bar{\mathcal{G}}(\tau-\tau')\bar{\mathcal{L}}^+ E(\tau')\right]\bar{\rho}(t_0)$$

$$= \mathcal{F}^{-1}\left\{\mathcal{F}\left(\int_{t_0}^t d\tau \bar{x}_i^\dagger \bar{\mathcal{G}}(t-\tau)\left[\bar{\mathcal{L}}^+ E(\tau)\int_{t_0}^\tau d\tau' \bar{\mathcal{G}}(\tau-\tau')\bar{\mathcal{L}}^- E^*(\tau') + \bar{\mathcal{L}}^- E^*(\tau)\int_{t_0}^\tau d\tau' \bar{\mathcal{G}}(\tau-\tau')\bar{\mathcal{L}}^+ E(\tau')\right]\bar{\rho}(t_0)\right)\right\}$$

$$= \mathcal{F}^{-1}\left\{\bar{x}_i^\dagger \bar{\mathcal{G}}(\omega)\left[\bar{\mathcal{L}}^+ \int d\omega' E(\omega-\omega')\bar{\mathcal{G}}(\omega')\bar{\mathcal{L}}^- E^*(\omega') + \bar{\mathcal{L}}^- \int d\omega' E^*(\omega-\omega')\bar{\mathcal{G}}(\omega')\bar{\mathcal{L}}^+ E(\omega')\right]\bar{\rho}(t_0)\right\}$$

$$= \mathcal{F}^{-1}\left\{\bar{x}_i^\dagger \bar{\mathcal{G}}(\omega)\mathcal{F}\left(|E(t)|^2\right)\right\}\left[\bar{\mathcal{L}}^+ \bar{\mathcal{G}}(-\omega_0)\bar{\mathcal{L}}^- + \bar{\mathcal{L}}^- \bar{\mathcal{G}}(\omega_0)\bar{\mathcal{L}}^+\right]\bar{\rho}(t_0)$$

$$= i0_+ \bar{x}_i^\dagger \bar{\mathcal{G}}(0)\left[\bar{\mathcal{L}}^+ \bar{\mathcal{G}}(-\omega_0)\bar{\mathcal{L}}^- + \bar{\mathcal{L}}^- \bar{\mathcal{G}}(\omega_0)\bar{\mathcal{L}}^+\right]\int_{t_0}^t d\tau\, |E(\tau)|^2\, \bar{\rho}(t_0)$$

$$\Pi_i(\infty) = i0_+ \bar{x}_i^\dagger \bar{\mathcal{G}}(0)\left[\bar{\mathcal{L}}^+ \bar{\mathcal{G}}(-\omega_0)\bar{\mathcal{L}}^- + \bar{\mathcal{L}}^- \bar{\mathcal{G}}(\omega_0)\bar{\mathcal{L}}^+\right]\bar{\rho}(t_0).$$

Assuming a conventional vectorization procedure (such that $\hat{a}\hat{b}\hat{c} \to \hat{a} \otimes \hat{c}^T \text{vec}(\hat{b})$), then we can write

$$\bar{\mathcal{G}}(\omega) = [i\omega_+ - \bar{A}]^{-1}$$

$$\bar{A} = \hat{a} \otimes \hat{1} + \hat{1} \otimes \hat{a}^* + \hat{L} \otimes \hat{L}^* + \sum_i \hat{Y}_i \otimes \hat{Y}_i^* + \sum_i \hat{X}_i \otimes \hat{X}_i^*$$

$$\hat{a} = -i\hat{H} - \frac{1}{2}\left[\hat{L}^\dagger L + \sum_i \hat{Y}^\dagger{}_i \hat{Y}_i + \sum_i \hat{X}_i^\dagger \hat{X}_i\right]$$

$$\bar{\mathcal{L}}^+ = \hat{L}^\dagger \otimes \hat{1}.$$

We then have

$$\Pi_i(\infty) = i0_+(\langle x_i| \otimes \langle x_i|)\bar{\mathcal{G}}(0_+))\left[(\hat{L}^\dagger \otimes \hat{1})\bar{\mathcal{G}}(-\omega_0)(\hat{1} \otimes \hat{L}^T) + (\hat{1} \otimes \hat{L}^T)\bar{\mathcal{G}}(\omega_0)(\hat{L}^\dagger \otimes \hat{1})\right](|0\rangle \otimes |0\rangle).$$

When the system is initialized in the ground state $|0\rangle$ and $\bar{\mathcal{L}}^\pm$ only couple the excited states to the ground state then

$$\bar{\mathcal{G}}(\omega_0)(\hat{L}^\dagger \otimes \hat{1})(|0\rangle \otimes |0\rangle) = ([i\omega_{0+} - \hat{a}]^{-1}\hat{L}^\dagger |0\rangle \otimes |0\rangle).$$

This can be demonstrated by expanding $\bar{\mathcal{G}}$ as a sum and observing that $\hat{a}\ket{0} = 0$. Defining $\bar{\mathcal{G}}_a(\omega) = [i\omega_+ - \hat{a}\otimes\hat{1} - \hat{1}\otimes\hat{a}^*]^{-1}$, $\hat{g}_a(\omega) = [i\omega_+ - \hat{a}]^{-1}$, and $\bar{A}_1 = \hat{L}\otimes\hat{L}^* + \sum_i \hat{Y}_i\otimes\hat{Y}_i^* + \sum_i \hat{X}_i\otimes\hat{X}_i$, we then have

$$\Pi_i(\infty) = i0_+(\bra{x_i}\otimes\bra{x_i})[1-\bar{\mathcal{G}}_a(0)\bar{A}_1]^{-1}\bar{\mathcal{G}}_a(0)\left[(\hat{L}\ket{0})\otimes(\hat{g}_a^\dagger(\omega_0)\hat{L}^T\ket{0}) + (\hat{g}_a(\omega_0)\hat{L}\ket{0})\otimes(\hat{L}^T\ket{0})\right]$$

$$= i0_+(\bra{x_i}\otimes\bra{x_i})[1-\bar{\mathcal{G}}_a(0)\bar{A}_1]^{-1}\bar{\mathcal{G}}_a(0)\left[\hat{1}\otimes\hat{g}_a^\dagger(\omega_0) + \hat{g}_a(\omega_0)\otimes\hat{1}\right](\hat{L}^\dagger\ket{0}\otimes\hat{L}^T\ket{0})$$

$$= i0_+(\bra{x_i}\otimes\bra{x_i})[1-\bar{\mathcal{G}}_a(0)\bar{A}_1]^{-1}\bar{\mathcal{G}}_a(0)\left[-\hat{1}\otimes\hat{a}^* - \hat{a}\otimes\hat{1}\right]\left(\hat{g}_a(\omega_0)\otimes\hat{g}_a^\dagger(\omega_0)\right)(\hat{L}^\dagger\ket{0}\otimes\hat{L}^T\ket{0})$$

$$= i0_+(\bra{x_i}\otimes\bra{x_i})[1-\bar{\mathcal{G}}_a(0)\bar{A}_1]^{-1}\left(\hat{g}_a(\omega_0)\otimes\hat{g}_a^\dagger(\omega_0)\right)(\hat{L}^\dagger\ket{0}\otimes\hat{L}^T\ket{0}). \tag{S1}$$

If $\hat{L}\hat{Y}_i = 0$ for all $i$ [conditions (1) and (2) given in the main text], then, upon expanding $[1-\bar{\mathcal{G}}_a(0)\bar{A}_1]^{-1}$ the only surviving term is

$$\Pi_i(\infty) = i0_+(\bra{x_i}\otimes\bra{x_i})\bar{\mathcal{G}}_a(0)(\sum_j \hat{Y}_j\otimes\hat{Y}_j^*)\left(\hat{g}_a(\omega_0)\otimes[-i\omega_{0+}-\hat{a}]^{-1}\right)(\hat{L}^\dagger\ket{0}\otimes\hat{L}^T\ket{0}).$$

Since the amplified states are stable, $(\bra{x_i}\otimes\bra{x_i})\bar{\mathcal{G}}_a(0)(\ket{x_i}\otimes\ket{x_i}) = \frac{1}{i0_+}$, and

$$\Pi_i(\infty) = \sum_j \left|\bra{x_i}\hat{Y}_j\hat{g}_a(\omega_0)\hat{L}^\dagger\ket{0}\right|^2$$

$$P_1 = \sum_{ij} \left|\bra{x_i}\hat{Y}_j\hat{g}_a(\omega_0)\hat{L}^\dagger\ket{0}\right|^2.$$

For simplicity, we have also assumed $\hat{Y}_j\hat{Y}_i = 0$ for all $i,j$, though this is not strictly necessary. Due to the above condition $\hat{L}\hat{X}_i = 0$, and since $\hat{L}^\dagger\hat{L} = \hat{L}^\dagger\ket{0}\bra{0}\hat{L}$,

$$\vec{Y}\hat{g}_a(\omega_0)\hat{L}^\dagger\ket{0} = \vec{Y}\left[i(\omega_{0+}-\hat{H}) + \frac{1}{2}\begin{pmatrix}\vec{Y}^\dagger & \hat{L}^\dagger\ket{0}\end{pmatrix}\begin{pmatrix}\vec{Y} \\ \bra{0}\hat{L}\end{pmatrix}\right]^{-1}\hat{L}^\dagger\ket{0}$$

where we used the shorthand $\vec{Y} = \begin{bmatrix}\hat{Y}_0 & \hat{Y}_1 & \cdots\end{bmatrix}$. Using the Woodbury identity, we can write

$$\vec{Y}\hat{g}_a(\omega_0)\hat{L}^\dagger\ket{0} = \vec{Y}\hat{g}_0(\omega_0)\hat{L}^\dagger\ket{0} - \vec{Y}\hat{g}_0(\omega_0)\begin{pmatrix}\vec{Y}^\dagger & \hat{L}^\dagger\ket{0}\end{pmatrix}\left[2 + \begin{pmatrix}\vec{Y} \\ \bra{0}\hat{L}\end{pmatrix}\hat{g}_0(\omega_0)\begin{pmatrix}\vec{Y}^\dagger & \hat{L}^\dagger\ket{0}\end{pmatrix}\right]^{-1}\begin{pmatrix}\vec{Y} \\ \bra{0}\hat{L}\end{pmatrix}\hat{g}_0(\omega_0)\hat{L}^\dagger\ket{0}$$

$$= \vec{Y}\hat{g}_0(\omega_0)\hat{L}^\dagger\ket{0} - \vec{Y}\hat{g}_0(\omega_0)\begin{pmatrix}\vec{Y}^\dagger & \hat{L}^\dagger\ket{0}\end{pmatrix}\begin{bmatrix}2 + \vec{Y}\hat{g}_0(\omega_0)\vec{Y}^\dagger & \vec{Y}\hat{g}_0(\omega_0)\hat{L}^\dagger\ket{0} \\ \bra{0}\hat{L}\hat{g}_0(\omega_0)\vec{Y}^\dagger & 2 + \bra{0}\hat{L}\hat{g}_0(\omega_0)\hat{L}^\dagger\ket{0}\end{bmatrix}^{-1}\begin{pmatrix}\vec{Y} \\ \bra{0}\hat{L}\end{pmatrix}\hat{g}_0(\omega_0)\hat{L}^\dagger\ket{0}$$

where we have set $\hat{g}_0(\omega_0) = \left[i(\omega_{0+} - \hat{H})\right]^{-1}$. Performing the inverse blockwise and simplifying, we obtain

$$\vec{Y}\hat{g}_a(\omega_0)\hat{L}^\dagger |0\rangle = \frac{4\left[2 + \vec{Y}\hat{g}_0(\omega_0)\vec{Y}^\dagger\right]^{-1} \vec{Y}\hat{g}_0(\omega_0)\hat{L}^\dagger |0\rangle}{2 + \langle 0|\hat{L}\left(\hat{g}_0(\omega_0) - \hat{g}_0(\omega_0)\vec{Y}^\dagger\left[2 + \vec{Y}\hat{g}_0(\omega_0)\vec{Y}^\dagger\right]^{-1}\vec{Y}\hat{g}_0(\omega_0)\right)\hat{L}^\dagger |0\rangle}$$

$$= \frac{\vec{Y}\left[\hat{g}_0(\omega_0)^{-1} + \frac{1}{2}\vec{Y}^\dagger\vec{Y}\right]^{-1}\hat{L}^\dagger |0\rangle}{1 + \frac{1}{2}\langle 0|\hat{L}\left[\hat{g}_0(\omega_0)^{-1} + \frac{1}{2}\vec{Y}^\dagger\vec{Y}\right]^{-1}\hat{L}^\dagger |0\rangle}.$$

Since $\vec{Y}^\dagger\vec{Y} = \sum_{ij} \hat{Y}_j^\dagger |x_i\rangle \langle x_i| \hat{Y}_j$ we can write

$$P_1 = \sum_{ij} \left|\langle x_i|\hat{Y}_j\hat{g}_a(\omega_0)\hat{L}^\dagger |0\rangle\right|^2$$

$$= \frac{\langle 0|\hat{L}\left[\hat{g}_0^\dagger(\omega_0)^{-1} + \frac{1}{2}\vec{Y}^\dagger\vec{Y}\right]^{-1}\vec{Y}^\dagger\vec{Y}\left[\hat{g}_0(\omega_0)^{-1} + \frac{1}{2}\vec{Y}^\dagger\vec{Y}\right]^{-1}\hat{L}^\dagger |0\rangle}{\left|1 + \frac{1}{2}\langle 0|\hat{L}\left[\hat{g}_0(\omega_0)^{-1} + \frac{1}{2}\vec{Y}^\dagger\vec{Y}\right]^{-1}\hat{L}^\dagger |0\rangle\right|^2}$$

$$= \frac{\langle 0|\hat{L}\hat{g}_0^\dagger(\omega_0)\vec{Y}^\dagger\left[1 + \frac{1}{4}\left|\vec{Y}\hat{g}_0(\omega_0)\vec{Y}^\dagger\right|^2\right]\vec{Y}\hat{g}_0(\omega_0)\hat{L}^\dagger |0\rangle}{\left|1 + \frac{1}{2}\langle 0|\hat{L}\hat{g}_0^\dagger(\omega_0)\vec{Y}^\dagger\left[1 + \frac{1}{4}\left|\vec{Y}\hat{g}_0(\omega_0)\vec{Y}^\dagger\right|^2\right]\vec{Y}\hat{g}_0(\omega_0)\hat{L}^\dagger |0\rangle + \frac{1}{4}\left|\langle 0|\hat{L}\left[\hat{g}_0(\omega_0)^{-1} + \frac{1}{2}\vec{Y}^\dagger\vec{Y}\right]^{-1}\hat{L}^\dagger |0\rangle\right|^2\right|^2}.$$

If $P_1 = 1$ then

$$\langle 0|\hat{L}\hat{g}_0^\dagger(\omega_0)\vec{Y}^\dagger\left[1 + \frac{1}{4}\left|\vec{Y}\hat{g}_0(\omega_0)\vec{Y}^\dagger\right|^2\right]\vec{Y}\hat{g}_0(\omega_0)\hat{L}^\dagger |0\rangle =$$

$$1 + \frac{1}{2}\langle 0|\hat{L}\hat{g}_0^\dagger(\omega_0)\vec{Y}^\dagger\left[1 + \frac{1}{4}\left|\vec{Y}\hat{g}_0(\omega_0)\vec{Y}^\dagger\right|^2\right]\vec{Y}\hat{g}_0(\omega_0)\hat{L}^\dagger |0\rangle + \frac{1}{2}\left|\langle 0|\hat{L}\left[\hat{g}_0(\omega_0)^{-1} + \frac{1}{4}\vec{Y}^\dagger\vec{Y}\right]^{-1}\hat{L}^\dagger |0\rangle\right|^2$$

$$\left|1 - \frac{1}{2}\langle 0|\hat{L}\left[\hat{g}_0(\omega_0)^{-1} + \frac{1}{2}\vec{Y}^\dagger\vec{Y}\right]^{-1}\hat{L}^\dagger |0\rangle\right|^2 = 0$$

$$\langle 0|\hat{L}\left[\hat{g}_0(\omega_0)^{-1} + \frac{1}{2}\vec{Y}^\dagger\vec{Y}\right]^{-1}\hat{L}^\dagger |0\rangle = 2.$$

Thus, provided that the LHS is real, the equality can be satisfied by scaling the operator $\hat{L}$, which can be achieved for the system in Fig. 3A by modulating $n_D$. Expanding the inverse yields

$$\langle 0|\hat{L}\hat{g}_0(\omega_0)^{-1}\hat{L}^\dagger |0\rangle + \sum_{i}^{i=1}\left(\frac{-1}{2}\langle 0|\hat{L}\hat{g}_0(\omega_0)^{-1}\vec{Y}^\dagger\vec{Y}\hat{g}_0(\omega_0)^{-1}\hat{L}^\dagger |0\rangle\right) = 2$$

$$i\langle 0|\hat{L}(\omega_0 - \hat{H})\hat{L}^\dagger |0\rangle + \sum_{i}^{i=1}\left(\frac{1}{2}\langle 0|\hat{L}(\omega_0 - \hat{H})\vec{Y}^\dagger\vec{Y}(\omega_0 - \hat{H})\hat{L}^\dagger |0\rangle\right) = 2.$$

We can see that the overall expression is real if the first vanishes, which will occur if the distribution describing the

system overall optical coupling spectrum is symmetric with respect to $\omega_0$.

### 3. TIGHT BINDING MODEL OF EXCITONIC SYSTEM

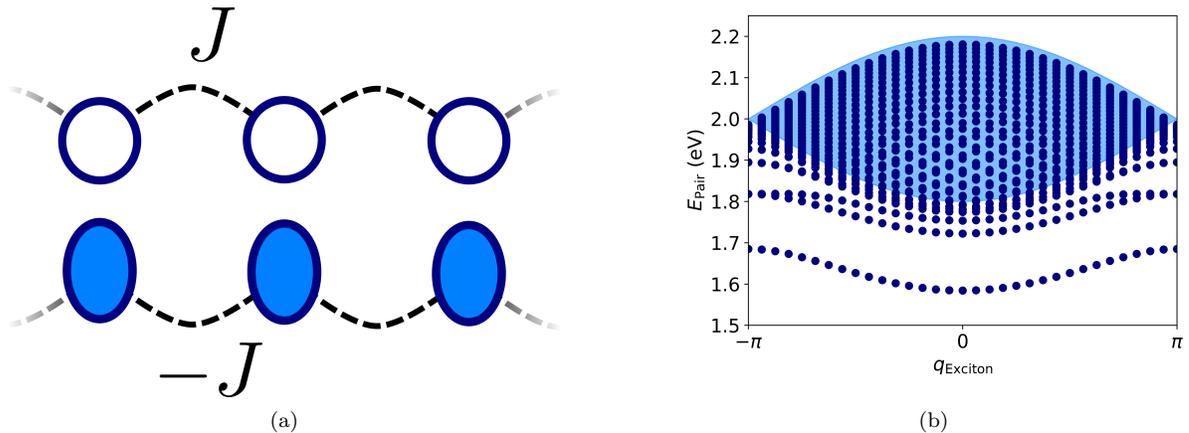

(a) (b)

FIG. S1. (a) A tight binding model with two states per site (e.g., $p_x$ and $p_z$ orbitals) with one filled and ione unfilled. Electron states couple to one another with strength $-J$ and hole states couple with strength $J$. Calculations were performed for 32 sites and periodic boundary conditions. (b) The e-h pair band structure, indexed by pair center-of-mass momentum $q_{\text{Exciton}}$. The shaded region shows the region occupied by the pair states in the absence of Coulomb interactions.

Following the procedure in Ref. [2, 3], we construct our system of identical sites with periodic boundary conditions Fig S1(a). Each site has a filled state (electron) and and unfilled state (hole), which are created by operators $d^\dagger$ and $c^\dagger$, respectively. We model Coulombic interactions using the Ohno potential. This gives the Hamiltonian

$$\hat{H} = \epsilon \sum_i (c^\dagger_i c_i - d^\dagger_i d_i) + J \sum_i (c^\dagger_i c_{i+1} + c^\dagger_{i+1} c_i - d^\dagger_i d_{i+1} - d^\dagger_{i+1} d_i)$$
$$+ \frac{1}{2} \sum_{ij} (c^\dagger_i c_i - d^\dagger_i d_i) \frac{U}{\sqrt{1 + (i-j)^2 \alpha^2}} (c^\dagger_j c_j - d^\dagger_j d_j).$$

We solve the many body eigenvalue equation in the subspace of single electron-hole pair states ($\psi_{ij} = c^\dagger_i d^\dagger_j |0\rangle$) directly to generate pair eigenstates. Due to the periodic boundary conditions, the center of mass momentum is a quantum number that can be used to index the states.

We consider a single, local, dispersionless phonon mode of the Holstein form, with carrier coupling

$$\hat{H}_{ph} = g_{ph} \sum_i \left(c^\dagger_i c_i - d^\dagger_i d_i\right) \left(b_i + b_i^\dagger\right)$$

$$\hat{Y}^x_{(ij)(ij)} = g(d_j c_i c^\dagger_i d^\dagger_j \delta_{xi} - d_j c_i c^\dagger_i d^\dagger_j \delta_{xj})$$

and field coupling

$$\hat{L} = \sum_i \gamma d^\dagger{}_i c_i. \tag{S2}$$

The field modes are incorporated into the model as indicated in the main text and the phonon modes as a thermalized bath within the Markov approximation [2]. We assume amplification of the excitonic ground state, which is essentially dark in the present model.

We convert the bath and light couplings to the pair eigenstate basis and evaluate the final efficiency using Equation (S1).

---


\end{document}